\documentclass[electronic]{vgtc}             

\ifpdf
  \pdfoutput=1\relax                   
  \usepackage{graphicx}                
  \DeclareGraphicsExtensions{.pdf,.png,.jpg,.jpeg} 
\else
  \usepackage{graphicx}                
  \DeclareGraphicsExtensions{.eps}     
\fi%

\graphicspath{{figures/}{pictures/}{images/}{./}} 

\usepackage{wrapfig}
\usepackage{microtype}                 
\PassOptionsToPackage{warn}{textcomp}  
\usepackage{textcomp}                  
\usepackage{mathptmx}                  
\usepackage{times}                     
\usepackage{cite}                      
\usepackage{tabu}                      
\usepackage{booktabs}                  
\usepackage{awesomebox}
\usepackage{float}
\usepackage{subfig}
\usepackage{physics,amsmath}
\usepackage{amssymb}
\usepackage{xcolor}
\usepackage{makecell}

\usepackage{multirow}

\usepackage[linesnumbered,ruled,vlined]{algorithm2e}

\SetCommentSty{commfont}

\vgtcinsertpkg

\title{GPU accelerated surface-based gaze mapping for XR experiences}

\author{Charles Javerliat\thanks{e-mail: charles.javerliat@enise.fr}\\ %
        \scriptsize Univ Lyon, Centrale Lyon, CNRS, INSA Lyon, UCBL, LIRIS, UMR5205 F-69130 Ecully France %
\and Guillaume Lavoué\thanks{e-mail: guillaume.lavoue@enise.fr}\\ %
     \scriptsize Univ Lyon, Centrale Lyon, CNRS, INSA Lyon, UCBL, LIRIS, UMR5205 F-69130 Ecully France %
}
        
\teaser{
  \centering
  \includegraphics[width=0.48\linewidth]{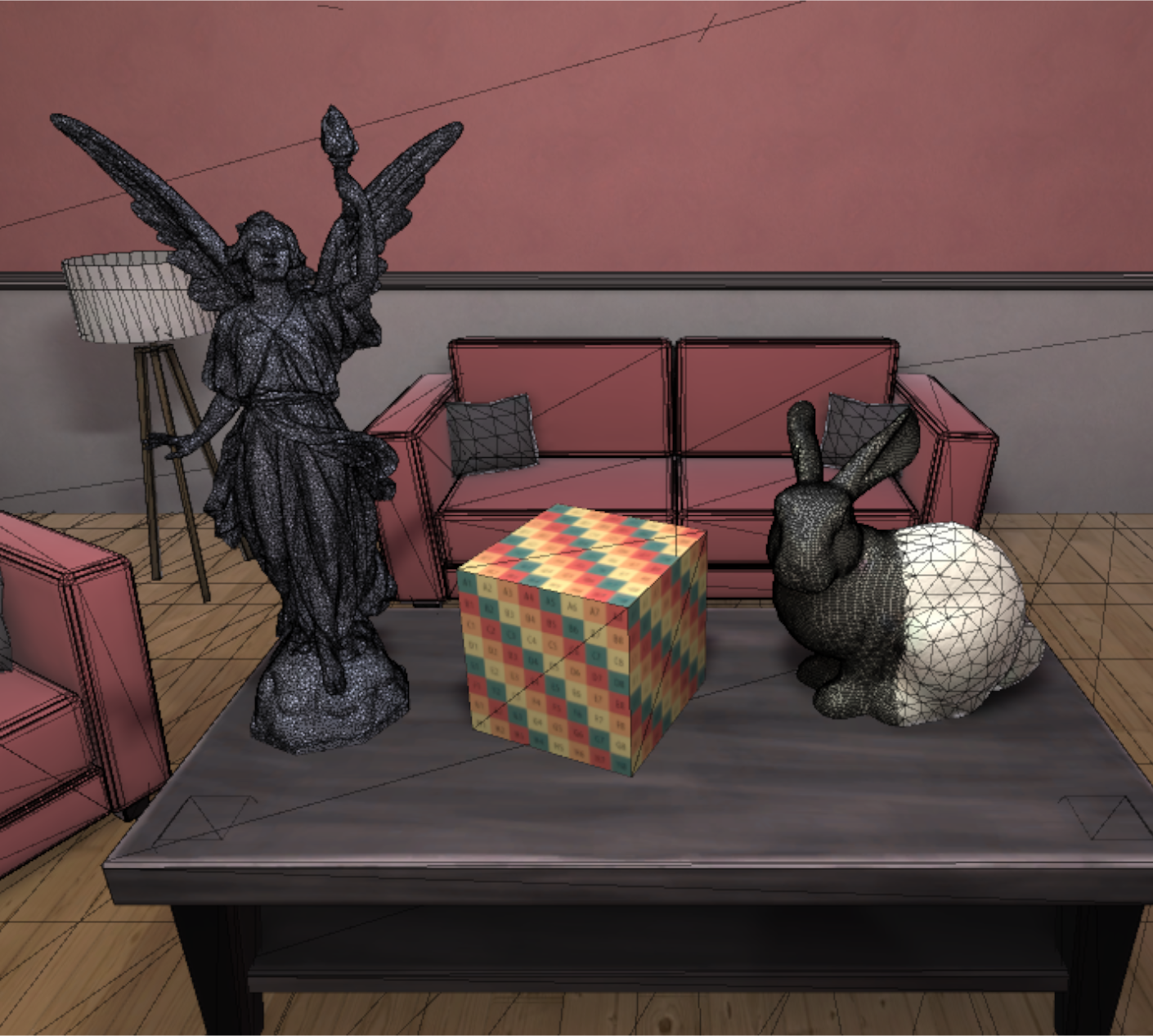}
  \;
  \includegraphics[width=0.48\linewidth]{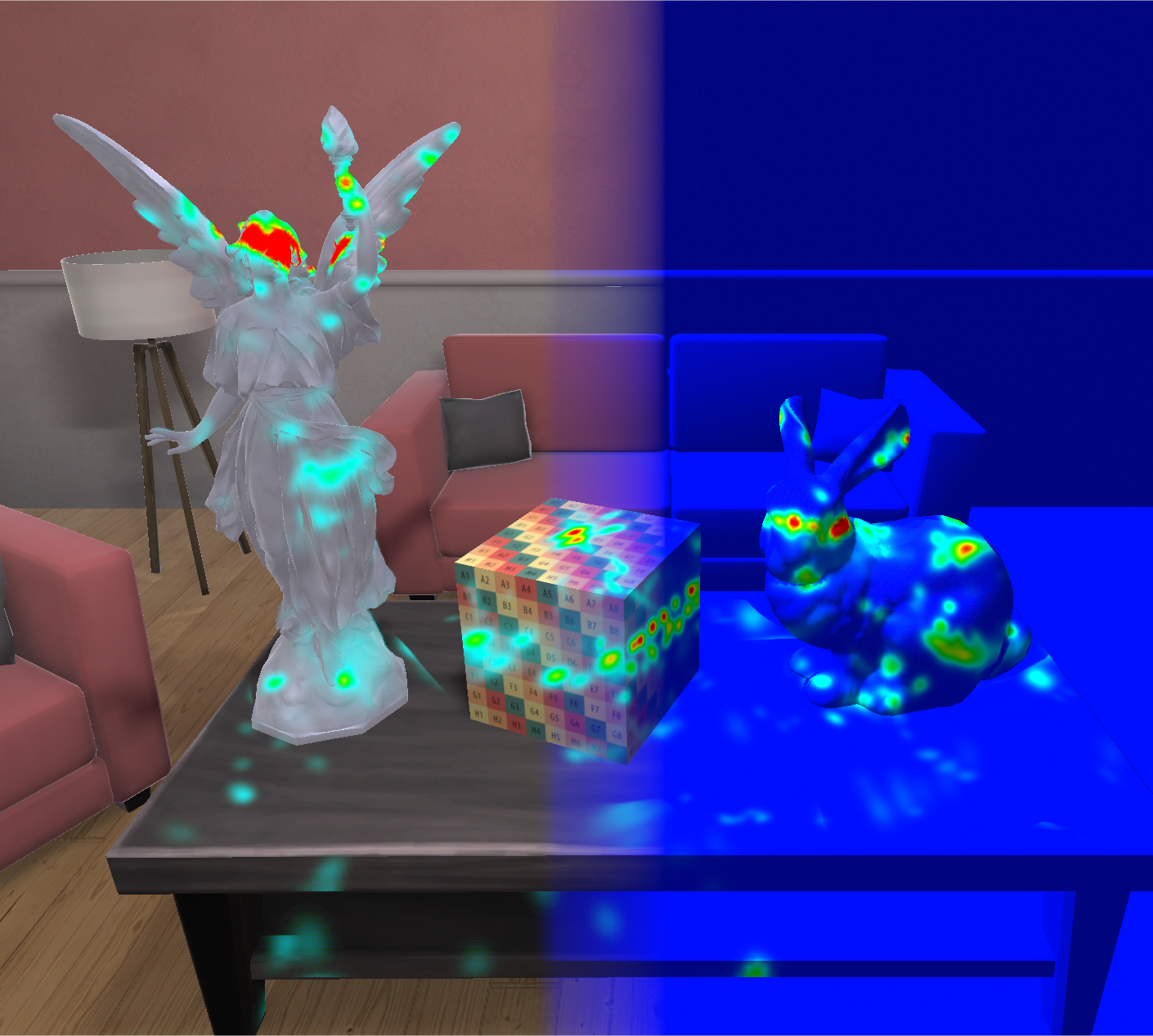}
  \caption{Generated fixation density map from a collection of 6000 eye-gaze fixations, recorded using a Vive Pro Eye HMD. The scene contains 800k triangles and a total of 2M fixation density map samples were generated in 6.9 seconds on the surface of the 4 meshes of interest. The objects have different challenging properties: the angel statue has no UV mapping, the cube has overlapping UVs, the bunny owns two very different mesh densities. Our algorithm can generate surface-based density maps agnostic to these properties.}
  \label{fig:teaser}
}

\abstract{
Extended reality is a fast-growing domain for which there is an increasing need to analyze and understand user behavior. In particular, understanding human visual attention during immersive experiences is crucial for many applications. The visualization and analysis of visual attention are commonly done by building fixation density maps from eye-tracking data. Such visual attention mapping is well mastered for 3 degrees of freedom (3DoF) experiences (\textit{i.e.}, involving 360 images or videos) but much less so for 6DoFs data, when the user can move freely in the 3D space. In that case, the visual attention information has to be mapped onto the 3D objects themselves. Some solutions exist for constructing such surface-based 6DoFs attention maps, however, they own several drawbacks: processing time, strong dependence on mesh resolution and/or texture mapping, and/or unpractical data representation for further processing. In this context, we propose a novel GPU-based algorithm that resolves the issues above while being generated in interactive time and rendered in real-time. Experiment on a challenging scene demonstrates the accuracy and robustness of our approach. To stimulate research in this area, the source code is publicly released and integrated into PLUME for ease of use in XR experiments.
} 

\CCScatlist{
  \CCScatTwelve{visualization}{eye-tracking}{immersive environments}{extended reality};
}

\begin{document}


\firstsection{Introduction}

\maketitle


Visual attention is a key point for understanding user behaviour and attention in XR experiments. Several recent works made important contributions in understanding and predicting visual attention for 3 degrees of freedom experiences involving 360$^{\circ}$ images and/or video \cite{Sitzmann_Serrano_Pavel_Agrawala_Gutierrez_Masia_Wetzstein_2017}\cite{Maranes_Gutierrez_Serrano_2020}\cite{Martin_Serrano_Bergman_Wetzstein_Masia_2021}\cite{Hu_Bulling_Li_Wang_2021}.
In those works, the visualization and analysis of visual attention are commonly done by building fixation density maps from raw eye-tracking data. In the case of 360$^{\circ}$ images and videos those heatmaps are represented in the 2D latitude/longitude plane.
Building such heat maps for 6 DoF immersive experiences involving 3D objects is not trivial since the attention information has to be mapped on the objects themselves (\textit{i.e.}, on their surface). Such object-based gaze mapping is of great interest for user behavior analysis and also to serve as ground truth for developing/learning visual attention prediction algorithms. Such attention prediction algorithms (at the object and surface levels) have a wide range of applications: optimizing storytelling and game balancing \cite{Koulieris_Drettakis_Cunningham_Mania_2015}, improving rendering performance (e.g., using Variable Rate Shading) by focusing the computations on the areas of interest from the user's perspective, fine-granularity foveated rendering without need of run-time eye-tracking, 3D scene optimization and so on.

Several solutions were proposed in the litterature for constructing 6DoFs attention maps\cite{Ding_Chen_2020}\cite{Alexiou_Xu_Ebrahimi_2019}\cite{Pfeiffer_Memili_2016}. The most advanced papers propose real-time generation and rendering, like Pfeiffer and Memili's\cite{Pfeiffer_Memili_2016}, but also present weaknesses such as strong dependence on the mesh resolution and/or texture mapping and unpractical data representation for using generated maps as ground-truth for prediction algorithms. Moreover, to our knowledge, there is no publicly available source code.


In this context, we propose a new interactive time generation and real-time rendered fixation density map algorithm, whose accuracy is configurable (e.g., in samples per square meter) and independent of the resolution of the mesh and the texture mapping; to stimulate research in this area, the algorithm is integrated in PLUME~\cite{javerliat_plume_2024} and the source code is publicly released in the PLUME Viewer repository\footnote{https://github.com/liris-xr/PLUME-Viewer} for ease of use in XR experiments. Our contributions are as follows:

\begin{itemize}
    \itemsep0em
    \item A new GPU-accelerated algorithm for generating surface-based fixation density maps in interactive time, for which the accuracy is independent of mesh resolution and UV mapping.
    \item A configurable generation where accuracy parameters can be changed.
    \item A filtering of the samples allowing a reduction of the generation time by a factor of 3.
    \item A real-time rendering of the previously generated density map with the ability to apply transformations to its values in real-time (\textit{i.e.} heatmap scale and associated colors).
    \item An open-source release of the code and its integration in PLUME.
\end{itemize}

\section{Related work}

Previous works on 6DoFs fixation density map include various techniques to map the data onto the 3D scene. In the paragraphs below we classify them according to the type of representation they consider.
We recall that for 2D images, fixation density maps are created by applying Gaussian convolutions on fixation points. This process aims at modeling the dispersion of the gaze around the point of fixation (due to the size of the fovea and the imprecision of the tracking device).

\subsection{Volume based}

Pfeiffer et al. proposed to represent the fixation density maps by volumes\cite{Pfeiffer_2012}, by dividing the 3D space into a voxel grid, storing the fixation values in each cell, and applying a 3D convolution. The 3D density map is rendered in real-time using a volumetric shader.
Although this method is pretty straightforward to implement, it presents weaknesses. The density map is bound to world space, when an object moves previous fixation points that landed on it will remain in a space left empty, leading to a more complex analysis of dynamic scenes. Moreover, a volumetric density map doesn't fully respect occlusion: a cell of the grid could contain a visible and hidden face of the mesh, but both would be included in the rendered volume. Moreover, the volumetric representation is not very intuitive in the sense that gaze ray naturally lands on surfaces rather than inside a volume.

\subsection{Object based}

Object-based representation associates each fixation point to a scene object\cite{Stellmach_Nacke_Dachselt_2010}. Each object stores a single scalar counting the total duration of fixations that landed on its mesh. This method has a macroscopic granularity, in the sense that it carries information about which objects are fixated the most, but without details on the area of fixation on the surface of the mesh. For example, Bernard et al. \cite{Bernhard_Stavrakis_Wimmer_2010} created a pipeline to create an importance map for each object of a scene, later used to predict users' visual attention in the context of a first-person shooter game.
It is particularly cheap resource-wise as it only requires one ray cast per fixation to determine the targeted object. It has the benefit of applying to dynamic scenes as the values are linked to objects and not to world space.

\subsection{Surface based}

Surface-based representation allows a fine granularity for fixation density maps. It allows to precisely represent which parts of a mesh were fixated. This is particularly useful to get information on which sections of a mesh attract the most attention. Surface-based maps are one of the most popular type of attention map in recent papers as it presents a lot of benefits for a deeper analysis of eye-tracking data. There are several surface-based representations that each have their pros and cons.

\subsubsection{Vertex-wise and triangle-wise}

When describing the surface of a mesh, several elements can be used to bound the attention map to it. The most straightforward method is relying directly on the definition of the mesh itself, using its vertices\cite{Stellmach_Nacke_Dachselt_2010}\cite{Lavoue_Cordier_Seo_Larabi_2018}\cite{Ding_Chen_2020} or triangles to store the value of the density map. For example, a single float per-vertex or triangle can be added and used inside of a shader to display the heatmap on the mesh. It is relatively lightweight as the memory consumption grows in $O(n)$ relatively to the vertex/triangle count.
However, this method is strongly dependent on the sampling resolution of the mesh. For example, a single cube composed of 2 triangles per face cannot contain more than 8 values when using vertices and 12 values when using triangles. Moreover, vertices and triangles of meshes created by designers are often non-uniformly distributed. For example a character's face will have a higher vertex/triangle density than the rest of its body. As a consequence, the density of information varying throughout the mesh has a direct impact on the resolution and the precision of the fixation density map.

\subsubsection{Pixel-wise}

Maurus et al.\cite{Maurus_Hammer_Beyerer_2014} proposed a method to project a Gaussian from screen space onto surfaces to form the fixation density map, and added the use of z-buffers to determine occluded areas for each point of view. This technique has the advantage of being relatively fast as it doesn't require computing raycasts into the scene for occlusion testing. In their paper, the presented method stores the z-buffers generated for each point of view and compute in real-time the fixation density map value for each fragment rendered on the screen.

This method is independent of the mesh triangle density of the objects in the scene as opposed to the previous method, which makes it a good solution for collecting fine details on the areas of fixations on objects. However this process is expensive in term of GPU memory, as it needs to store a z-buffer as a texture for each point of view. The memory consumption makes it difficult to be applied in a mobile context, where the device memory is limited. Additionally, this paper presents a method that works well on static scenes (they took the example of a museum exhibition), but as the density map values are not bound to the scene objects, they loose their meaning when those are moved. Moreover, to our knowledge, the data can't be exported for external analysis. 

\subsubsection{Texel-wise}

\begin{figure}[ht]
    \centering
    {{\includegraphics[width=0.5\linewidth]{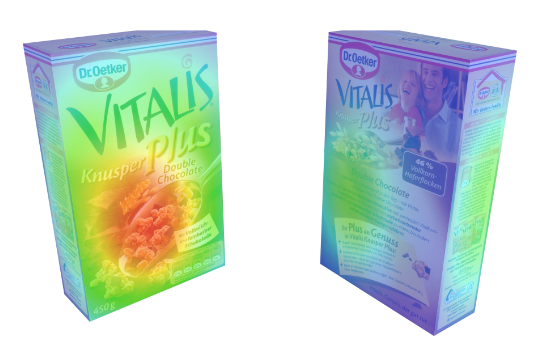} }}%
    \qquad
    {{\includegraphics[width=0.3\linewidth]{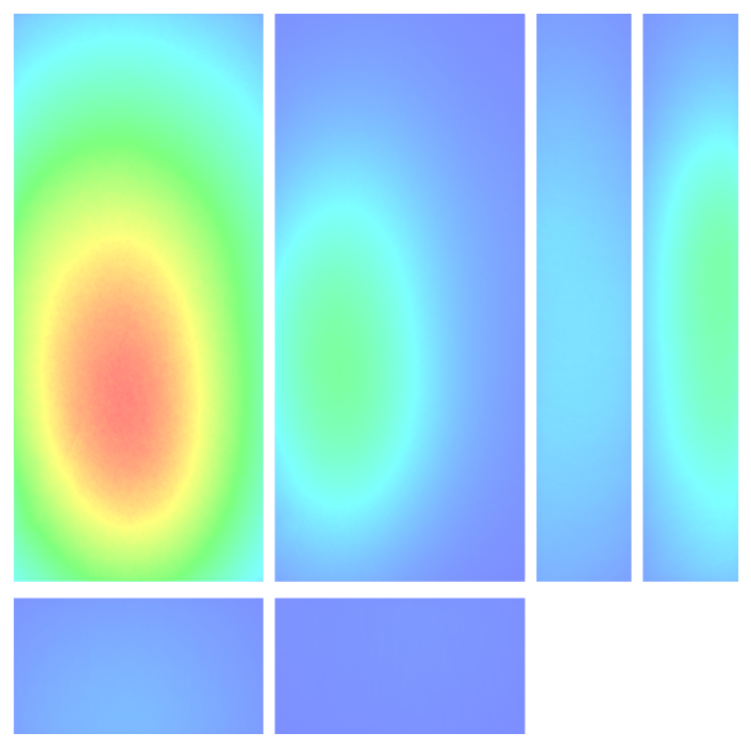} }}%
    \caption{Texel based 3D attention map and its associated texture image. (Reprinted from Pfeiffer et Memili\cite{Pfeiffer_Memili_2016})}
    \label{fig:cereal}
\end{figure}

To preserve the advantage of a detailed fixation density map without being too heavy memory-wise and to have values bound to separate objects of the scene, Pfeiffer et Memili \cite{Pfeiffer_Memili_2016} proposed a great method that stores the fixation density map values into a texture image (\textit{cf.} figure \ref{fig:cereal}). One texture is generated per object. Each value of the map is computed using a Gaussian distribution represented as a ball to prevent artefacts from projecting a 2D Gaussian into 3D space, and uses a similar method to Maurus et al.\cite{Maurus_Hammer_Beyerer_2014} for occlusion testing using a z-buffer.

This method is the most appropriate for real-time rendering of attention maps. It supports dynamic scenes, animated models, and require less memory compared to previous methods. However, it also comes with its own limitations. Most importantly, the attention map texture is strongly coupled with the way the UV mapping was laid out by model artists. This can cause problems when dealing with models with non-uniform texel density, leading to disparities in the distribution of information over the mesh. In the case of CAD models, UV mapping may not even exist. Moreover, the authors of the paper pointed out the issue of overlapping UV coordinates. In fact, artists can reuse parts of their textures on the surface of their mesh, resulting in false heatmap where values are shared by two distincts triangles of the mesh.
The texture method is fast and does not use as much memory as the pixel-wise method, but the texture in which the information is stored is not very convenient for external usage. For instance, one might need to use the generated fixation density map values as ground truth data for prediction algorithms, to find relations between objects shapes, materials, etc and visual attention. In this case, a ready-to-use data representation would be to have the values of the fixation density map associated to the mesh's triangles index and barycentric coordinates or 3D points in mesh's local space, rather than stored inside of a texture, loosing direct spatial context of where the value comes from the mesh, and would require an extra step to map the texture to mesh to extract spatial context. Additionally, the program proposed by the authors generates the fixation density map while the experiment is running. This allows for a real-time generation as they don't need to re-render the scene from each point of view post-experiment, but also comes with downsides. For instance, if you require a higher resolution fixation density map, you can't regenerate it without having to redo the experiment. You also can't generate the fixation density map with a finer control of the considered fixations, for example if you want to consider only fixations within a given time period.

To tackle those issues, we propose a method agnostic to UV mapping and mesh resolution. We propose a representation inspired by \emph{Mesh Colors} proposed by Yuksel et al.\cite{Yuksel2008} to store the map values, consisting of a sub-sampling of each triangle of the mesh. Our method propose a generation of the fixation density map in interactive time, after the experiment is done, and a real-time rendering. By using PLUME, we are able to record the state of the scene continuously and reproject the fixations post-experiment, allowing for a lossless recording. Our solution has the ability to export the data as a list of value associated to the triangle index and barycentric coordinates for each sample for easier external processing.

\section{Generation algorithm}

\subsection{Mesh points sampling}
\label{section:sampling}

We resolve issues related to UV mapping and mesh resolution by storing the attention map on points sampled on the surface. We use a quasi-uniform samples distribution based on a triangle sub-sampling method. We adapt the number of samples in triangles depending of their surface area to achieve an acceptable quasi-uniform distribution over all the mesh. The proposed representation is not as uniform as a Poisson disk sampling, but allows for a $O(1)$ indexing of samples and thus a good compromise between uniformity and computational cost.

\subsubsection{Samples indexing}

Our representation is inspired by the work of Yuksel et al.\cite{Yuksel2008}, who originally proposed a way to store colors inside mesh triangles using an equally subdivided barycentric space. We adapted the proposed method to better fit our need, particularly for performance reasons, such as indexing triangles samples in a $O(1)$ complexity, and storing a single float per sample instead of a full 4-channels color (RGBA) value. The subdivision of each triangle is parameterized by $r$, the resolution of the subdivision. It can be seen as the number of sub-triangles on the side of the original one. When subdividing a triangle with a resolution of $r$, the n\textsuperscript{th} triangle formula gives us the number of samples $\#p$ in the triangle as:

\begin{equation}
    \#p=f(r)=\frac{(r+1)(r+2)}{2}
    \label{samples_cardinality}
\end{equation}

The original Mesh Colors algorithm proposes a way to reference samples by their barycentric coordinates in the triangle. As we also needed to do the opposite, \textit{i.e.}, find a sample barycentric coordinates by its index, we developed a method that calculates the sample barycentric coordinates given its index ($idx$) in a $O(1)$ complexity. To find the $row$ of a given sample of index $idx$ we start by calculating the inverse function of $f$.

\[ f^{-1}(x)=\frac{-3\pm\sqrt{8x+1}}{2} \]

We are only interested in the positive value and the integer part corresponding to the $row$. Moreover, as our indexing system starts at 0 we need to add an offset of 1 to the $idx$ inside the formula, giving us the final expression for the $row$ that can be used to express the column index $col$:

\begin{equation}
 \begin{aligned}
  row&=\left\lceil \frac{-3 + \sqrt{8\cdot idx + 9}}{2} \right\rceil\\
  col&=idx-f(row-1)
 \end{aligned}
\end{equation}

Hence, the barycentric coordinates for our sample of index $idx$ in the triangles are:

\begin{equation}
    (w_1, w_2, w_3) = \begin{bmatrix}
        col / r \\
        1 - row / r \\
    	1 - \frac{col + row}{r}
    \end{bmatrix}
    \label{barycentric_coords}
\end{equation}

The pseudocode for the conversion from sample index to barycentric weights is provided in supplementary material.

\subsubsection{Adaptive resolution}

\begin{figure}[ht]
    \centering
    \subfloat[\centering Without adaptive resolution]{{\includegraphics[width=0.4\linewidth]{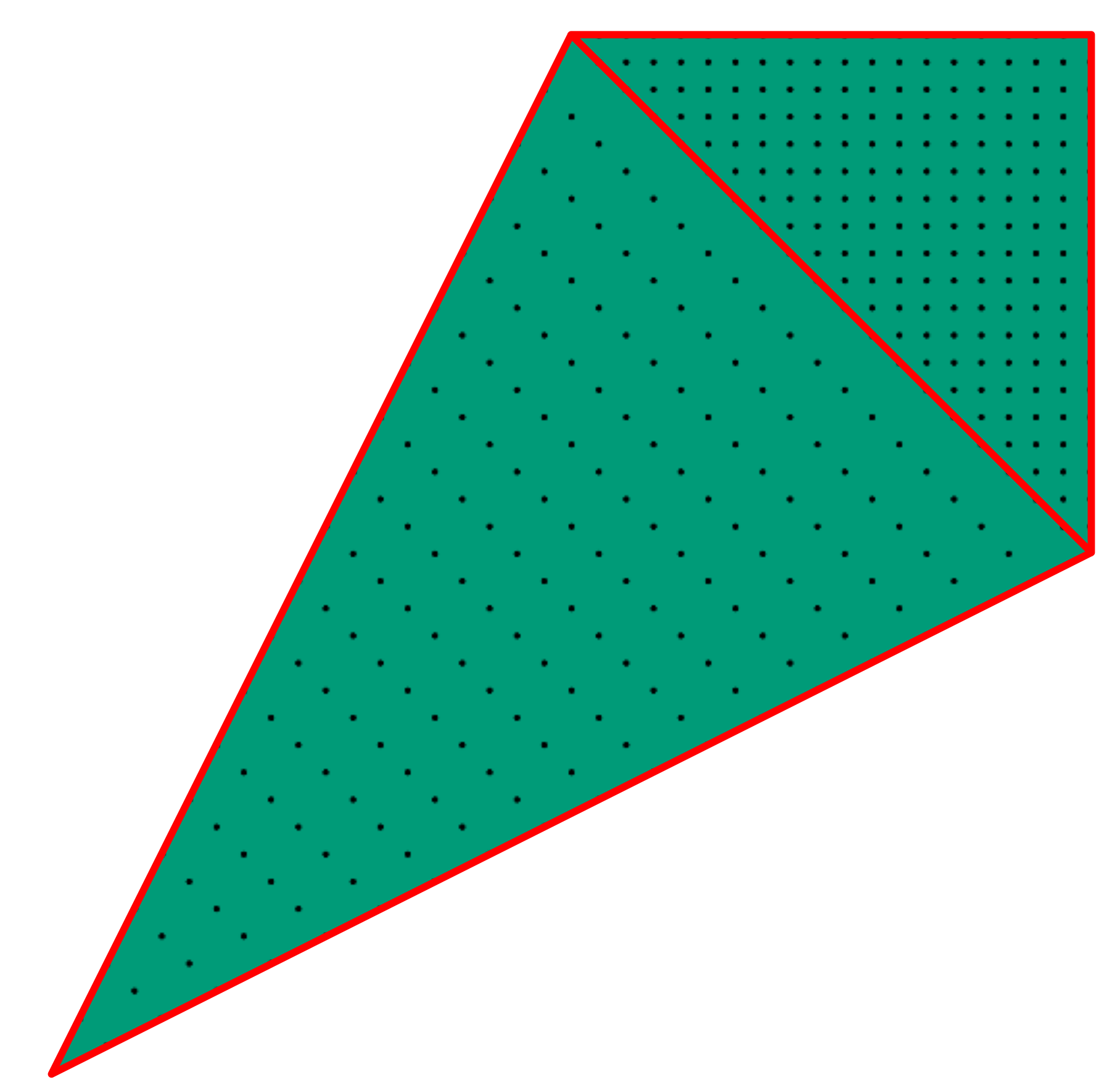} }}%
    \qquad
    \subfloat[\centering With adaptive resolution ]{{\includegraphics[width=0.4\linewidth]{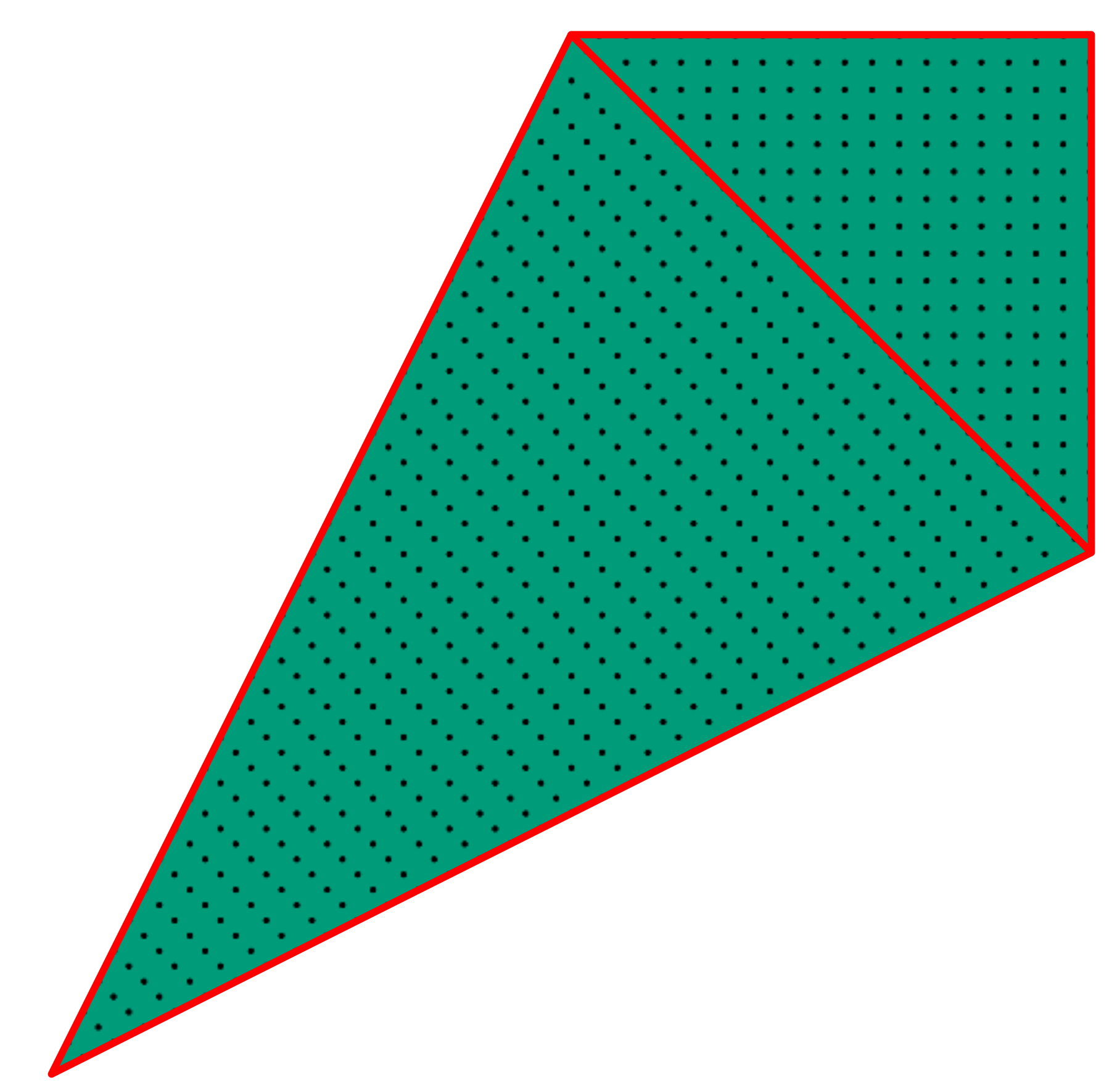} }}%
    \caption{Uniformity of samples distribution with and without adaptive resolution. The edges of the triangles are represented in red, the samples are represented as black dots. In the left image, samples are further apart from each other in the stretched triangle, this will be even more noticeable in models with larger triangles for low detailed areas. }
    \label{fig:samples-distribution-example}%
\end{figure}

The triangles of a mesh are rarely uniform in terms of scale across a model. For instance, it may have a very large amount of small triangles in high detailed areas, and fewer but larger triangles in low detailed sections. As a consequence, applying the same resolution $r$ to all the triangles of the mesh would result in an unbalanced distribution of samples.

We propose the following algorithm to adapt the sampling for each triangle in order to obtain a globally quasi-uniform distribution (see figure \ref{fig:samples-distribution-example}.b). Let $k$ be a parameter corresponding to the number of sample points per squared meter we want to achieve, $\mathcal{A}$ the area of the triangle and $r$ the resolution of the triangle. By using Heron's formula, we compute the area of the triangle. We start by calculating the length of each edge using the triangle's vertices $\{v_0, v_1, v_2\}$:

\begin{equation*}
a=||v_1-v_0||\qquad b=||v_2-v_1||\qquad c=||v_2-v_0||
\end{equation*}

Applying Heron's formula gives us the area of the triangle $\mathcal{A}$ depending on its semi-perimeter $s$:

\begin{equation*}
s = \frac{a+b+c}{2}\qquad
\mathcal{A} = \sqrt{s(s-a)(s-b)(s-c)}
\end{equation*}

We want $\frac{\#p}{\mathcal{A}}=k$, we combine this constraint with \eqref{samples_cardinality} and solve the resulting system of equations for $r$:
\begin{equation*}
\begin{aligned}
    \begin{split}
    &\;\begin{cases}
        \frac{\#p}{\mathcal{A}}=k\\
        \#p=\frac{(r+1)(r+2)}{2}
    \end{cases}
    \Leftrightarrow &\;
    r^2+3r+2-2k\mathcal{A}=0
    \end{split}
\end{aligned}
\end{equation*}

As $k > 0$ and $\mathcal{A} > 0$, hence $\Delta > 0$, we have two solutions in $\mathbb{R}$:

\begin{equation*}
\Delta=1+8k\mathcal{A}\qquad
r_{1,2}=\frac{-3\pm\sqrt{\Delta}}{2}
\end{equation*}

We want the resolution to be positive and greater or equal to 1 which is only true if $\Delta \geq 25$. When this is not satisfied, we clamp the resolution $r$ to 1 to ensure that each triangle has at least its three vertices storing a value. Finally $r$ is expressed as:

\begin{equation}
  r = \left\{\begin{alignedat}{2}
    & 1 && \Delta < 25 \\
    & \frac{-3+\sqrt{1+8k\mathcal{A}}}{2} \qquad && \Delta \ge 25
  \end{alignedat}\right.
\end{equation}

When applied, the adaptive resolution results in good quasi-uniform distribution of sample points on triangles of different size as shown in the figure \ref{fig:samples-distribution-example}.

The triangle subdivision we use ensures that the samples are evenly spaced across the surface of an equilateral triangle. In practice, triangles are rarely equilaterals in a model, but using this representation allows a cheap way of indexing samples in a $O(1)$ complexity, as demonstrated above, and is good enough for achieving an overall quasi-uniform distribution of samples on the mesh. Additionally, when a mesh resolution becomes too high (with very small triangles) the sampling becomes linked again to the number of triangles as we can't set a sampling resolution smaller than the mesh resolution (each triangle contains at least 3 samples: one per vertex). However this issue can be counterbalanced by increasing the sampling resolution to put more samples per square meter. We consider it to be a good compromise between distribution uniformity and computing cost.

\subsection{Fixation density map generation}

The generation of the fixation density map is composed of several steps. We start by accumulating projected Gaussian distributions corresponding to  gaze rays with their associated angular deviation from each point of view (see figure \ref{gaussian-projection-principle}). This step takes into account any occlusion that could occur when looking at an object. In a final pass, we normalize the values so that each sample's value is included in $[0, 1]$. Those steps are detailed below.

\subsubsection{Gaussian projection onto mesh}

\begin{figure}[ht]
    \centering
    \includegraphics[width=0.8\linewidth]{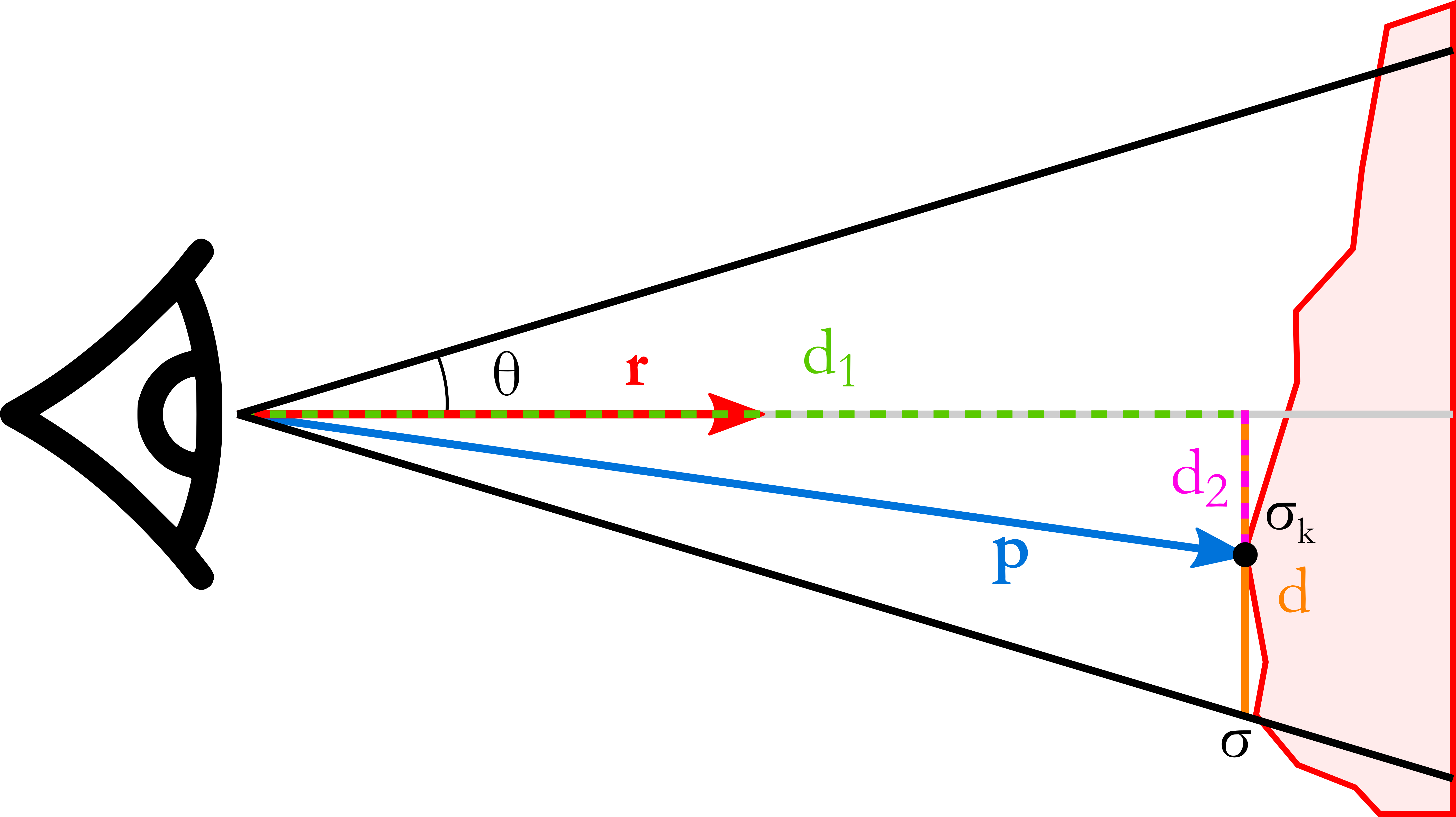}
    \caption{Gaussian projection principle. $\vb{r}$ is the gaze ray direction, $\vb{p}$ is a vector going from the camera to a considered sample point. The cone represents the deviation of the gaze ray with $\sigma$ corresponding to the angular deviation $\theta$. Note that the Gaussian does also propagate outside of the cone represented on this figure. In our case, we propagate the values up to $4\sigma$, corresponding to approximately 100\% of the Gaussian distribution.}
    \label{gaussian-projection-principle}
\end{figure}

A simple ray is not enough to accurately represent the human fixation. The eyes do not fixate on a point, but rather on an area with a size corresponding to the fovea's. Additionally, the eye tracking device has a limited precision. This imprecision on the measured gaze ray angle and visual dispersion has to be taken into account. For this, it is common to model the gaze ray as cone, representing the projection of a Gaussian distribution in 3D space.

Let $\vb{p}$ be a vector going from the camera to a considered sample point in space and $\vb{r}$ the gaze ray direction. Both are expressed in the camera local space.
We note $d_1$ the scalar projection and $d_2$ the scalar rejection of $\vb{p}$ onto $\vb{r}$ (see figure \ref{gaussian-projection-principle}).

\begin{equation*}
d_1=\vb{p}\cdot \frac{\vb{r}}{\norm{\vb{r}}}\qquad
d_2=\vb{p}\cdot \frac{\vb{r}^\perp}{\norm{\vb{r}}}
\end{equation*}

We want to model the Gaussian distribution so that the angular deviation $\theta$ corresponds to one standard deviation.
Let $\sigma$ be the standard deviation of the Gaussian distribution and $d$ be the cone radius at the distance $d_1$ from the camera origin:

\begin{equation*}
\sigma = tan \theta\qquad
d = d_1 \cdot tan \theta
\end{equation*}

We define $\sigma_k$ as the product of the standard deviation $\sigma$ and a factor $k$ corresponding to the ratio $\frac{d_2}{d}$:

\begin{equation*}
  \sigma_k = \left\{\begin{alignedat}{2}
    & 0 && d = 0 \\
    & \frac{d_2}{d} \cdot \sigma \qquad && d > 0
  \end{alignedat}\right.
\end{equation*}

For each triangle sample we can compute the associated Gaussian value for a given gaze ray. To aggregate the contribution of each fixation, we weight them by their duration.

\[ g(\vb{p}, \vb{r}, t)=\frac{t}{\sigma \cdot \sqrt{2\pi}} \cdot \exp{\frac{-\sigma_k^2}{2\sigma^2}} \]

The pseudocode for the Gaussian value computation algorithm is provided in supplementary material.

\subsubsection{Occlusion support}

\begin{figure}[ht]
    \centering
    \includegraphics[width=\linewidth]{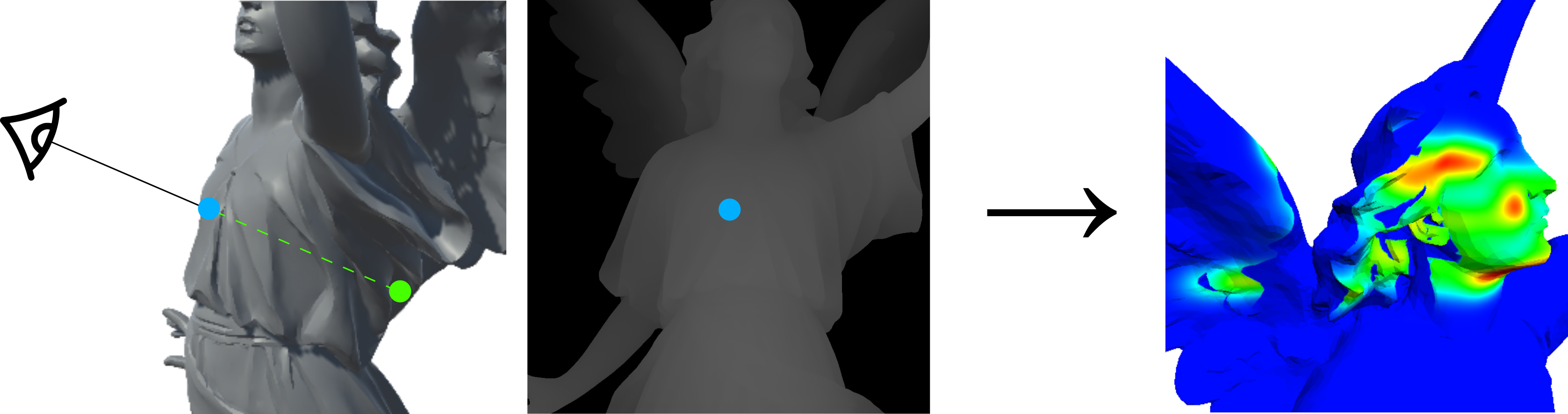}
    
    \caption{The blue sample point on the front of the model is closer to the camera than the green one in the back. In the z-buffer (right image), the depth of the closest point (blue) is stored in the pixel corresponding to its projection in screen space. The smaller the depth value, the whiter the pixel. When performing the occlusion check for the green sample, its depth is different from the one stored inside its corresponding pixel inside the z-buffer ($\pm\epsilon$), meaning that the green point is not visible. An example of multiple fixations projected onto the mesh with occlusion support can be seen on the right image. }
    \label{z-buffer}
\end{figure}

When looking at an object, some of it is not always visible depending on the point of view. Therefore, the computation of the Gaussian values should only be performed on visible samples.
The easiest and fastest way to perform an occlusion check is to use a z-buffer as previously cited papers used before\cite{Maurus_Hammer_Beyerer_2014}\cite{Pfeiffer_Memili_2016}. This buffer, generated on highly specialized hardware parts of GPUs and using the painter's algorithm, stores the minimal depth of 3D points projected on the screen. When two 3D points are projected onto the same buffer pixel, the shortest depth is kept. To know if a sample is visible, we compare its depth with the closest depth stored in the z-buffer. If the values are close enough ($\pm \epsilon$), the sample is visible from the point of view the z-buffer was generated for. An example of an occlusion-check is depicted in figure \ref{z-buffer}.

\subsubsection{Normalization}

A final pass is used to normalize the values after the aggregation. Normalization is useful for filtering or applying transformations on the values later on. For example to tweak the render of the heatmap by changing the color scale, enhancing contrast, etc. To normalize the values, we keep track of the maximum value attributed to the sample points during the aggregation. For each sample, we can then divide its value by this maximum, resulting in a normalized fixation density map with each value included in the interval $[0, 1]$.

\subsection{Optimizations}


\subsubsection{GPU acceleration}

\begin{figure}[ht]
    \centering
    \includegraphics[width=0.8\linewidth]{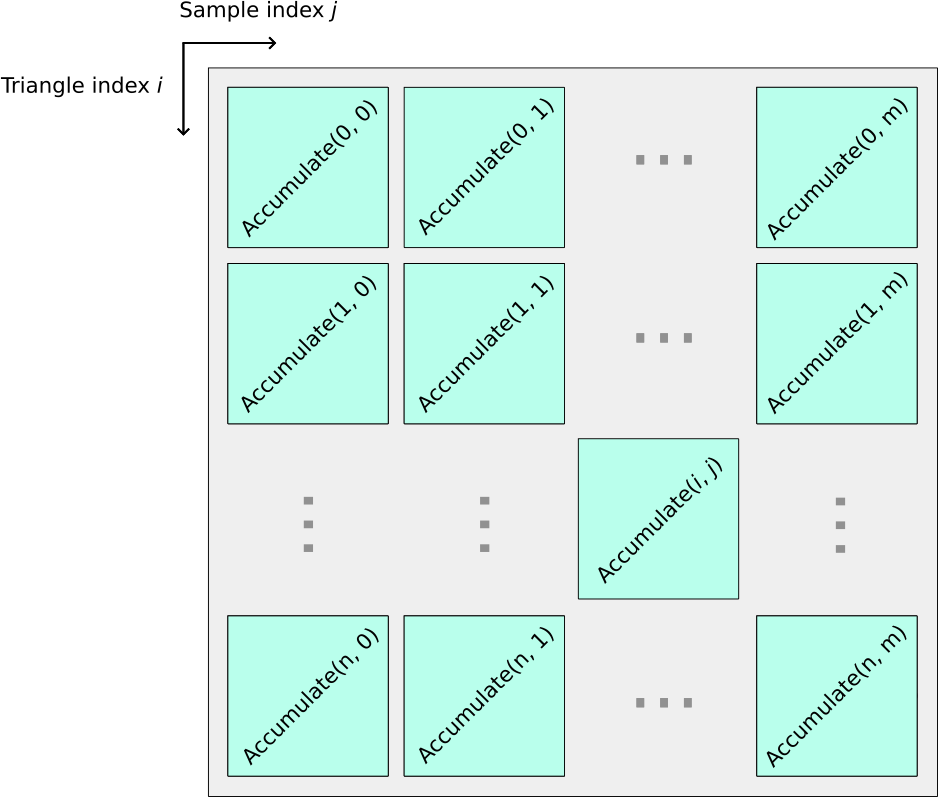}
    \caption{The accumulation kernel can be run separately for each sample. By using the GPU, this allows for a fast generation of the fixation density map. The same is done for the normalization kernel.}
    \label{fig:kernel-parallel}
\end{figure}

Thanks to the samples indexing system and the fact that the Gaussian values can be computed independently for each sample, this make our algorithm easily parallelizable on a GPU. We can create one GPU kernel responsible for accumulating values, and one for normalizing them. As each triangle of index $i$ contains several samples of index $j$, we can run the accumulation and normalization kernels for every sample in parallel (see figure \ref{fig:kernel-parallel}).

For the normalization, to avoid adding an extra pass going through all the samples to find the maximum value, we keep track of it inside the accumulation pass. However, as each thread can update the maximum value at any time, the latter is subject to a race condition. Using a mutex inside the accumulation kernel locking the writing operations on the maximum value, we ensure that no race condition can occur in this critical section of the code.

\subsubsection{Sample filtering}

When accumulating values for a considered fixation, the naive approach is to generate a z-buffer for the full screen and go through every object in the scene, running the kernel for every sample contained in the mesh's triangles. It appears clearly that this method is greedy and doesn't scale well. This method is not applicable for complex scenes. Moreover, for a single fixation, we lost a lot of time and information rendering the z-buffer in parts of the screen where the gaze doesn't even land on. In fact, only a portion of the scene is included in the projection of the Gaussian where values aren't equals to zero. Knowing this, we can optimize the accumulation pass by reducing drastically the number of considered samples.


Approximately 100\% of the projected Gaussian non-zero values are included inside a cone corresponding to 4 standard deviation from the central gaze ray. We can use that to our advantage to filter samples, reduce the number of rendered triangles inside the z-buffer and concentrate the z-buffer pixels information in the section that is relevant for our algorithm.

\begin{figure*}[ht]
    \centering
    \includegraphics[width=\linewidth]{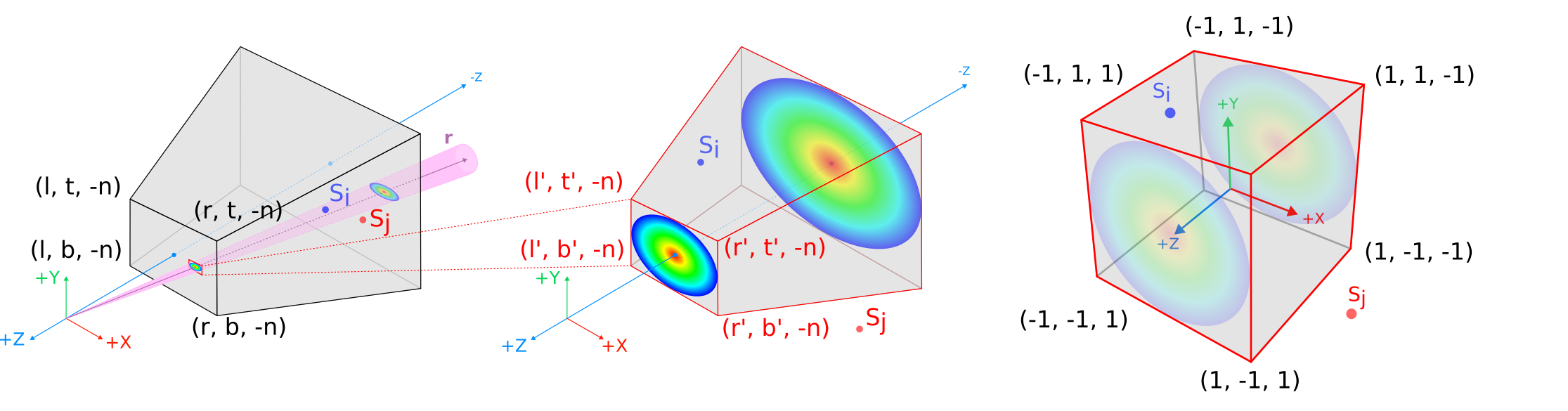}
    \caption{The original camera (on the left) can be cropped to fit the $4\sigma$ cone (in the middle). The resulting projection matrix preserves the original perspective. By normalizing the coordinates in the normalized device coordinates space, we can get rid of the samples outside of the cube (\textit{cf.} $S_j$ in the right figure).}
    \label{crop-mtx}
\end{figure*}

To implement this optimization, we need to crop the camera view to fit the $4\sigma$ cone (\textit{cf.} figure \ref{crop-mtx}). To obtain the corresponding projection matrix, we need to determine the new parameters $l'$, $r'$, $t'$, $b'$ based on the original projection matrix parameters $l$, $r$, $t$, $b$, $n$, $f$. Those parameters correspond respectively to the left, right, top, bottom, near and far parameters that describe how the view frustum is constructed (\textit{cf.} figure \ref{crop-mtx}).

\begin{wrapfigure}[11]{r}{0.2\textwidth}
    \centering
    \includegraphics[width=0.2\textwidth]{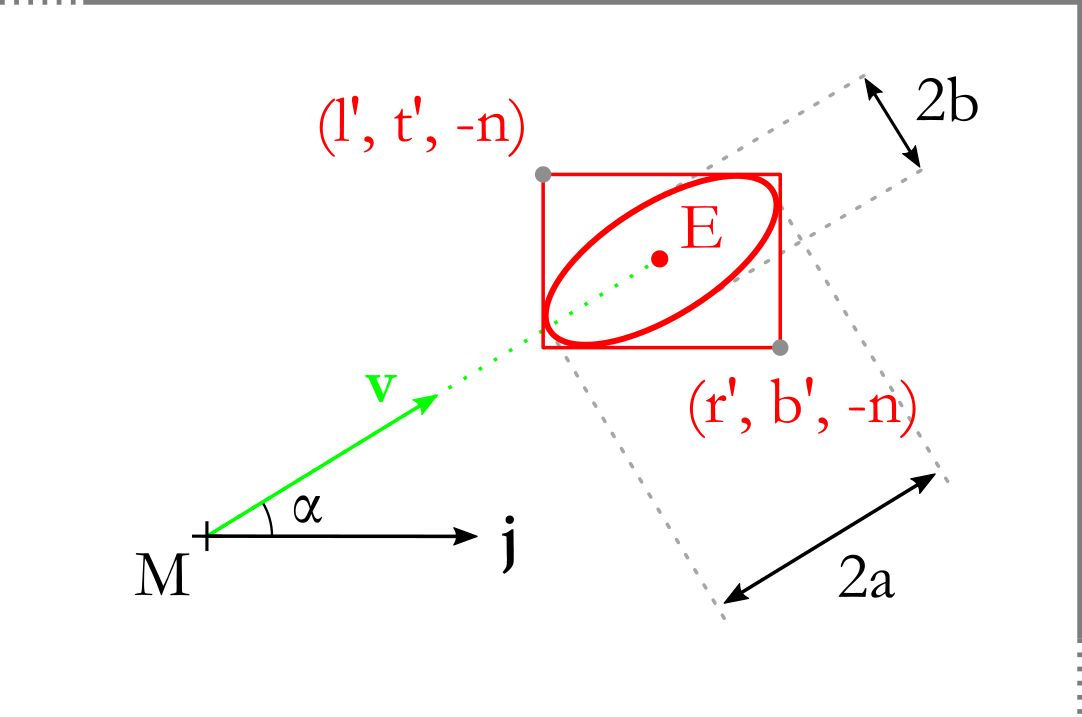}
    \vspace{-10pt}
    \caption{}
    \label{fig:params}
\end{wrapfigure}
For this, we need to find the bounding box of the ellipse corresponding to the intersection of the gaze ray cone and the near clip plane.


Let the geometric parameters of the intersection ellipse between the the near clip plane and the gaze ray cone be: $a$ its major radius, $b$ its minor radius, $\alpha$ its inclination and $E$ its center point (\textit{cf.} figure \ref{fig:params}). The bounds of the ellipse can be calculated as the following (calculation details are developed in appendix \ref{ellipse-bounds-appendix}):

\begin{equation}
\begin{aligned}
l'&=E_x - \sqrt{a^2 \cdot cos^2\alpha + b^2 \cdot sin^2\alpha}\\
r'&=E_x + \sqrt{a^2 \cdot cos^2\alpha + b^2 \cdot sin^2\alpha}\\
b'&=E_y - \sqrt{a^2 \cdot sin^2\alpha + b^2 \cdot cos^2\alpha}\\
t'&=E_y + \sqrt{a^2 \cdot sin^2\alpha + b^2 \cdot cos^2\alpha}
\end{aligned}
\label{ellipse-geometric-params-eq}
\end{equation}

Using \eqref{ellipse-geometric-params-eq}, the crop matrix is expressed as a perspective projection matrix with the new parameters:

\begin{equation*}
\begin{aligned}
\begin{bmatrix}
\frac{2 \cdot n}{r' - l'} & 0 & \frac{r' + l'}{r' - l'} & 0\\
0 & \frac{2\cdot n}{t'-b'} & \frac{t'+b'}{t'-b'} & 0\\
0 & 0 & \frac{-(f + n)}{f - n} & \frac{-2 \cdot f \cdot n}{f - n}\\
0 & 0 & -1 & 0
\end{bmatrix}
\end{aligned}
\end{equation*}

To filter the samples, we use this crop projection matrix to convert their coordinates into normalized device coordinates (see figure \ref{crop-mtx} on the right). We then filter samples outside of the NDC cube (interval $[-1, 1]$) as those are outside of the view frustum of the cropped camera. An example is given in figure \ref{crop-mtx} with $S_i$ being inside and $S_j$ being outside of the cropped camera view frustum. The pseudocode for the accumulation and normalization kernels are available in supplementary material.

\subsubsection{Export format}

In order to use the generated fixation density map outside of our program, we propose a convenient export format. For each sample, its value is exported as well as any contextual data such as the object and triangle it belongs to, its barycentric coordinates in the triangle, mesh local space coordinates and world space coordinates. This way, one can easily use the data as a ground truth map for prediction algorithms.

\subsubsection{XREcho and PLUME integration}

\begin{figure}[htbp]
    \centering
    \includegraphics[width=\linewidth]{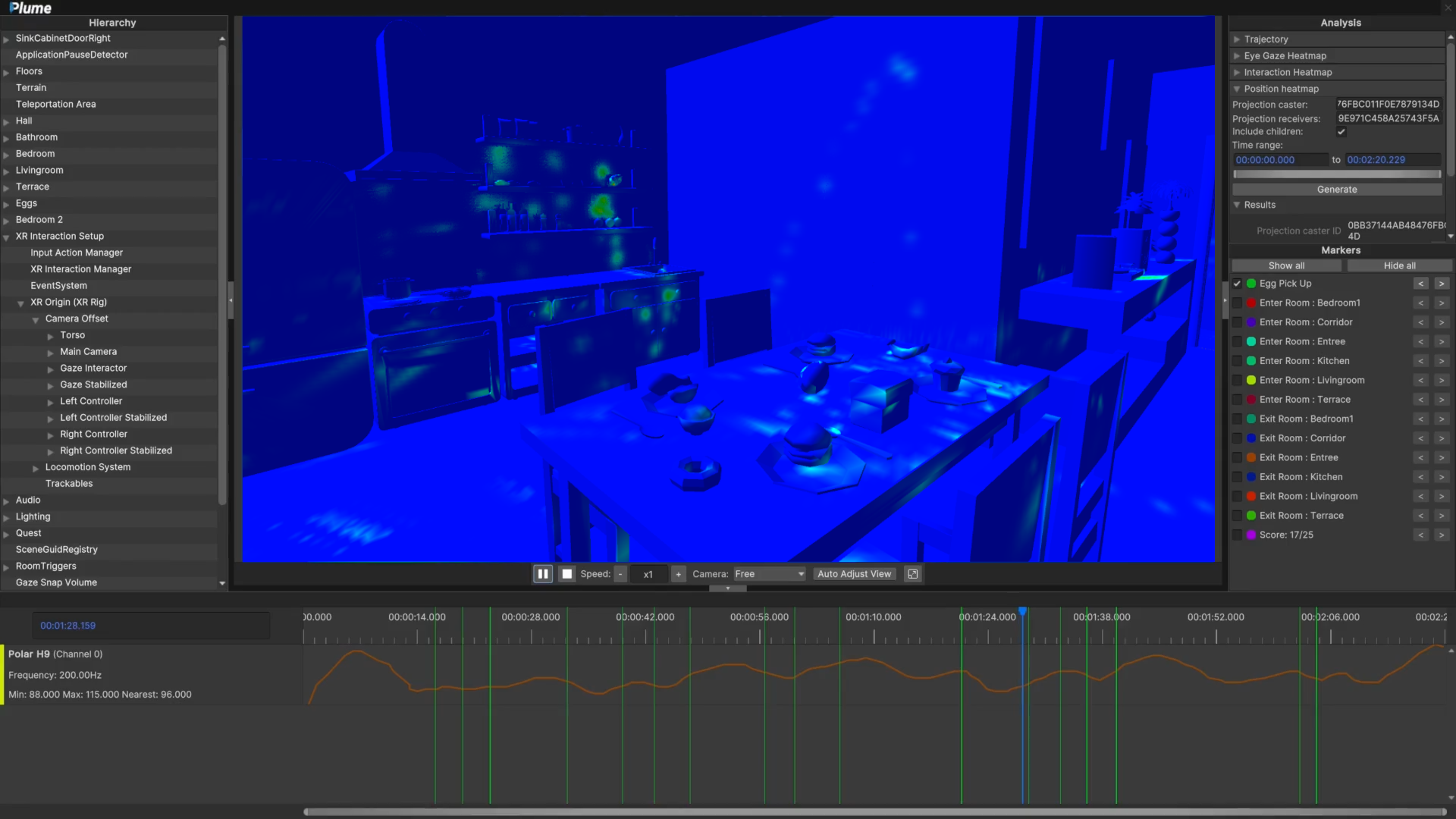}
    \caption{Visualization of eye gaze heatmap integration within the PLUME Viewer~\cite{javerliat_plume_2024}. The central panel displays the reconstructed three-dimensional scene at a specific time, with the computed eye gaze heatmap projected and rendered directly onto the surfaces of the scene objects.}
    \label{fig:heatmap_plume}
\end{figure}

Our heatmap generation and rendering algorithms are integrated into XREcho~\cite{Villenave_2022} and PLUME\cite{javerliat_plume_2024} (see Figure~\ref{fig:heatmap_plume}). Both allows to record, replay and visualize the user behavior during XR session. It also records the gaze information. Our tools thus enriches the analysis tools already present in these frameworks. Multiple parameters for the generation can be changed: resolution of the z-buffer, sampling resolution for different meshes, time interval in which we consider the fixations and angular deviation $\theta$ of the gaze ray cone. The scale of the heatmap can be changed in real-time, as well as the colors, thanks to a custom shader converting the fixation density map normalized values to RGB.

\section{Experimental results}

To demonstrate the robustness of our method, we ran the algorithm on a challenging scene with meshes that would cause artefacts with previous solutions (see figure \ref{fig:teaser}). It is composed of a mesh with no UV mapping (the angel statue), one with overlapping UVs (the cube) and one with unbalanced distribution of vertices (the rabbit). Note that we specifically filtered the samples to ignore projections on the walls and floor for better clarity on the rendering. Fixations were recorded during a single session of exploration of the scene, using a Vive Pro Eye HMD. A real-time rescale transformation was applied on normalized values to show larger red areas on the render. We generated the map with different parameters and listed the results in table \ref{tab:challenging-scene}. We repeated each generation 10 times and provide mean values, standard-deviations and confidence intervals. The fixation density map was generated in a few seconds, and the sample filtering optimization allows to reach a speed of generation approximately 3 times faster. We tested our algoritm with 3 different sampling resolution, one that we consider to be low (10000 samples/m$^2$), medium (40000 samples/m$^2$) and high (80000 samples/m$^2$).

\begin{table}[H]
\begin{tabular}{ ||c|c|c|c|| } 
\hline
$n=10$ & $\mu$ & $\sigma$ & CI (95\%) \\
\hline\hline
\makecell{With sample filtering\\10000 samples/m$^2$} & 6.58s & 0.27 & $[6.39s, 6.77s]$ \\
\hline
\makecell{With sample filtering\\40000 samples/m$^2$} & 6.90s & 0.31 & $[6.68s, 7.12s]$ \\ 
\hline
\makecell{With sample filtering\\80000 samples/m$^2$} & 11.0s & 0.60 & $[10.6s, 11.4s]$ \\ 
\hline
\makecell{Without sample filtering\\40000 samples/m$^2$} & 18.0s & 0.68 & $[17.5s, 18.5s]$ \\
\hline
\end{tabular}
\caption{Fixation density map generation time for 6000 fixations on the challenging scene depicted in figure \ref{fig:teaser} on a NVIDIA GeForce GTX 1060.}
\label{tab:challenging-scene}
\end{table}

\begin{figure}[H]
    \centering
    \includegraphics[width=\linewidth]{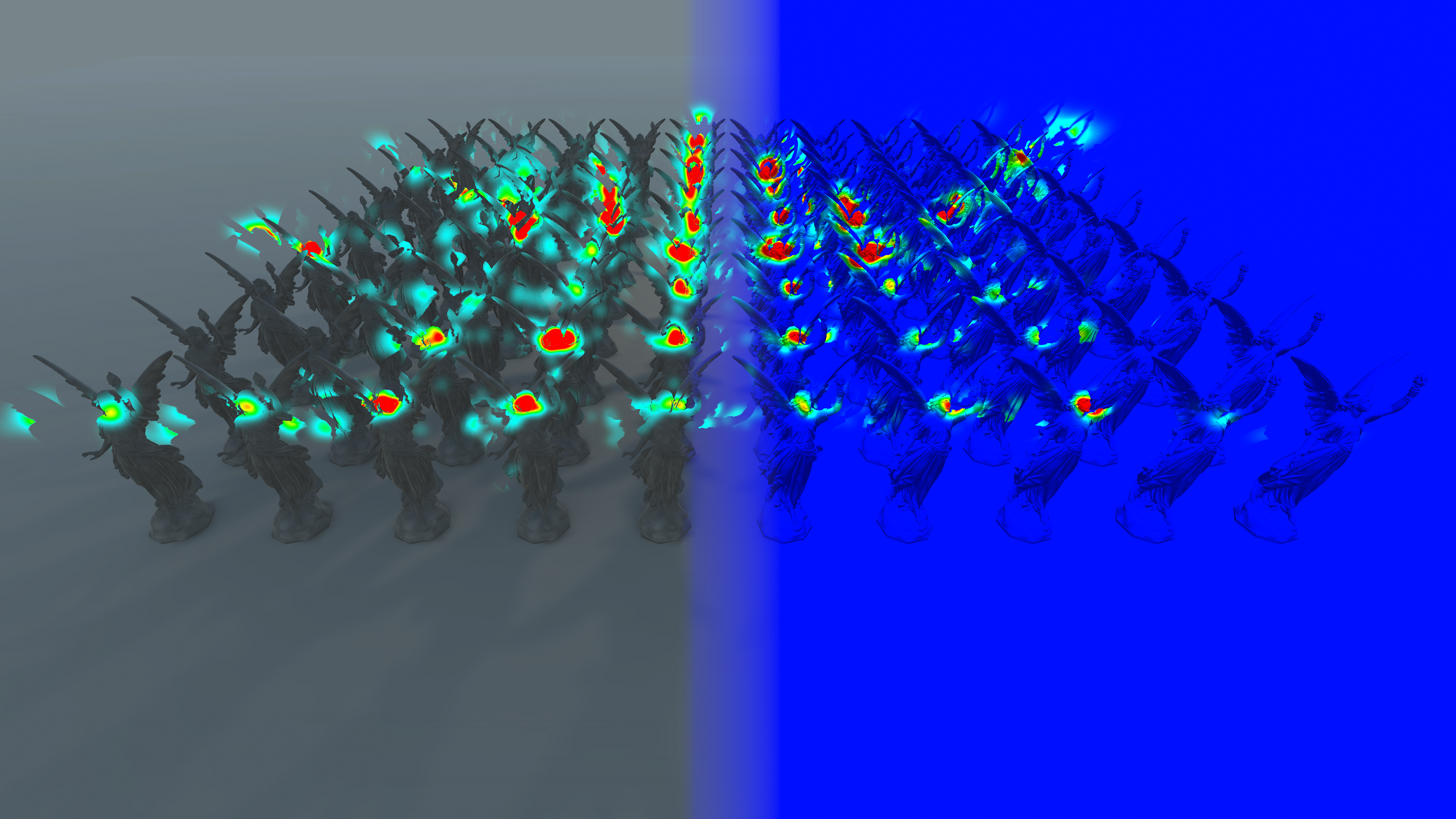}
    \caption{Stress test scene containing 100 angel statues and one large plane. The scene contains a total of 10M triangles. The 236M samples of the map were generated in under a minute.}
    \label{fig:stress-test}
\end{figure}

To test the scalability of our system, we also ran a stress test on a scene with a lot of highly detailed meshes. We placed 100 angel statues on one large plane. Each model has approximately 100k triangles with a sampling resolution of 80000 samples per square meter. The scene contains a total of 10M triangles and a total of 236M samples distributed over the meshes; the generation for 6000 fixations took under a minute. The render (real-time) is shown in figure \ref{fig:stress-test} and numerical results are listed in table \ref{tab:stress-test-scene}.

\begin{table}[H]
\begin{tabular}{ ||c|c|c|c|| } 
\hline
$n=10$ & $\mu$ & $\sigma$ & CI (95\%) \\
\hline\hline
\makecell{With sample filtering\\80000 samples/m$^2$} & 58.3s & 0.57 & $[57.9s, 58.7s]$ \\
\hline
\makecell{Without sample filtering\\80000 samples/m$^2$} & 156s & 1.53 & $[155s, 157s]$ \\
\hline
\end{tabular}
\caption{Fixation density map generation times for 6000 fixations on the stress test scene depicted in figure \ref{fig:stress-test} on a NVIDIA GeForce GTX 1060.}
\label{tab:stress-test-scene}
\end{table}

\section{Conclusion and perspectives}

In this paper, we presented an new open-source GPU-accelerated algorithm for generating surface-based fixation density map in an interactive time and rendering it in real-time. We show that our algorithm accuracy is independent of mesh resolution and UV mapping by executing it on a challenging scene that would present artefacts with previous state-of-the-art methods. The scalability of our representation was tested through a complex scene containing approximately 10M triangles with positive results: the fixation density map was generated under a minute. Future work includes the use of this tool to create large databases of 6DoF attention maps to feed learning algorithms for predicting eye gaze, or for improving the understanding of user behavior during immersive experiments.


\newpage

\section{Appendix}

\appendix

\section{Ellipse intersection between gaze ray cone and near clip plane}
\label{ellipse-bounds-appendix}

\begin{figure}[ht]
    \centering
    \includegraphics[width=\linewidth]{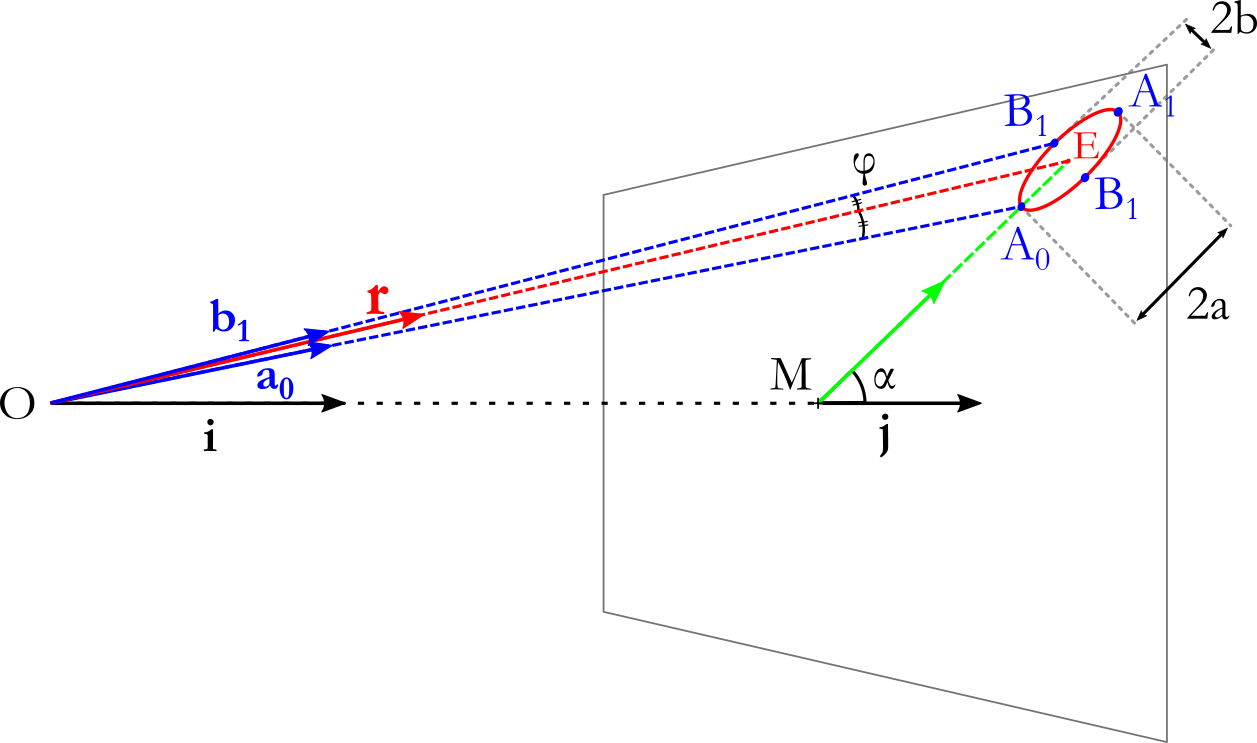}
    \caption{Gaze ray cone intersection with the near clip plane. O is the position of the point of view, M is the middle point of the near clip plane, E is the intersection of the gaze ray with the plane and corresponds to the center of the ellipse, $\vb{i}$ and $\vb{j}$ are two vectors of the camera base, respectively the forward and right vector.}
    \label{ellipse-intersection}
\end{figure}

To find the bounding box of the ellipse, corresponding to the intersection of the gaze ray cone with the near clip plane, we first calculate its geometric parameters: its center $E$, major radius $a$, shorter radius $b$ and inclination $\alpha$. We can find the position of $A_0$, $A_1$, $B_0$ and $B_1$ by rotating the gaze ray $\vb{r}$ at an angle $\varphi$, the angle corresponding to 4 standard deviation from the gaze ray, along the direction of the major axis and shorter axis. We know $\sigma = tan\theta$, therefore we can express $\varphi$ as:

\begin{equation*}
    \begin{cases}
      \sigma = tan\theta\\
      4\sigma=tan\varphi
    \end{cases}
    \;\Rightarrow\;
    \varphi=atan(4\sigma)
\end{equation*}

The rotation axes direction vectors $\vb{u_1}$ and $\vb{u_2}$ used to respectively rotate the ray $\vb{r}$ along the major and shorter axis are expressed as:

\begin{equation*}
\vb{u_1} = \vb{r} \cross \vb{i}\qquad
\vb{u_2} = \vb{r} \cross \vb{u_1}
\end{equation*}

To rotate the vector $\vb{r}$ by an angle $\varphi$ around an axis directed by a vector $\vb{u}$, we use a quaternion of the form:
\begin{equation*}
q=cos\frac{\varphi}{2}+\vb{u} \cdot sin\frac{\varphi}{2}
\end{equation*}
The resulting rotated vector $\vb{r'}$ is calculated using the formula:
\begin{equation*}
\vb{r'}=q\cdot\vb{r}\cdot q^{-1}=(cos\frac{\varphi}{2}+\vb{u} \cdot sin\frac{\varphi}{2})\cdot\vb{r}\cdot(cos\frac{\varphi}{2} - \vb{u} \cdot sin\frac{\varphi}{2})
\end{equation*}

Applying this method, we can calculate the vectors $\vb{a_0}$, $\vb{a_1}$, $\vb{b_0}$, $\vb{b_1}$ directing the lines going from $O$ to $A_0$, $A_1$, $B_0$ and $B_1$:

\begin{equation*}
\begin{aligned}
\vb{a_0} &= (cos\frac{-\varphi}{2}+\vb{u_1} \cdot sin\frac{-\varphi}{2})\cdot\vb{r}\cdot(cos\frac{-\varphi}{2} - \vb{u_1} \cdot sin\frac{-\varphi}{2})\\
\vb{a_1} &= (cos\frac{\varphi}{2}+\vb{u_1} \cdot sin\frac{\varphi}{2})\cdot\vb{r}\cdot(cos\frac{\varphi}{2} - \vb{u_1} \cdot sin\frac{\varphi}{2})\\
\vb{b_0} &= (cos\frac{-\varphi}{2}+\vb{u_2} \cdot sin\frac{-\varphi}{2})\cdot\vb{r}\cdot(cos\frac{-\varphi}{2} - \vb{u_2} \cdot sin\frac{-\varphi}{2})\\
\vb{b_1} &= (cos\frac{\varphi}{2}+\vb{u_2} \cdot sin\frac{\varphi}{2})\cdot\vb{r}\cdot(cos\frac{\varphi}{2} - \vb{u_2} \cdot sin\frac{\varphi}{2})\\
\end{aligned}
\end{equation*}

For any line directed by a vector $\vb{v}$ and going through point $O$, the intersection point with the near clip plane located at $z=-n$ can be found by solving the following system of equation:

\begin{equation}
\begin{aligned}
    \begin{cases}
    x = O_x + t \vb{v}_x\\
    y = O_y + t \vb{v}_y\\
    z = O_z + t \vb{v}_z = -n
    \end{cases}
    \Leftrightarrow
    \begin{cases}
    t = \frac{-n - O_z}{\vb{v}_z}\\
    x = O_x + \frac{-n - O_z}{\vb{v}_z} \vb{v}_x\\
    y = O_y + \frac{-n - O_z}{\vb{v}_z} \vb{v}_y\\
    z = -n
    \end{cases}
\end{aligned}
\label{eq:point-intersect}
\end{equation}

By applying (\ref{eq:point-intersect}) with $\vb{a_0}$, $\vb{a_1}$, $\vb{b_0}$, $\vb{b_1}$ and $O=(0, 0, 0)$, we can express $A_0$, $A_1$, $B_0$ and $B_1$ as:

\begin{equation*}
\begin{aligned}
A_0 &=
\begin{pmatrix}
\frac{-n}{a_{0z}}a_{0x}\\
\frac{-n}{a_{0z}}a_{0y}\\
-n
\end{pmatrix}\qquad
A_1 &=
\begin{pmatrix}
\frac{-n}{a_{1z}}a_{1x}\\
\frac{-n}{a_{1z}}a_{1y}\\
-n
\end{pmatrix}\\
B_0 &=
\begin{pmatrix}
\frac{-n}{b_{0z}}b_{0x}\\
\frac{-n}{b_{0z}}b_{0y}\\
-n
\end{pmatrix}\qquad
B_1 &=
\begin{pmatrix}
\frac{-n}{b_{1z}}b_{1x}\\
\frac{-n}{b_{1z}}b_{1y}\\
-n
\end{pmatrix}
\end{aligned}
\end{equation*}

The geometric parameters of our ellipse can be calculated:

\begin{equation*}
\begin{aligned}
E &=
\begin{pmatrix}
\frac{-n}{r_{z}}r_{x}\\
\frac{-n}{r_{z}}r_{y}\\
-n
\end{pmatrix}\qquad
M =
\begin{pmatrix}
0\\
0\\
-n
\end{pmatrix}\\ 
a&=\frac{||A_1-A_0||}{2}=\frac{n}{2}\sqrt{(\frac{a_{0x}}{a_{0z}}-\frac{a_{1x}}{a_{1z}})^2+(\frac{a_{0y}}{a_{0z}}-\frac{a_{1y}}{a_{1z}})^2}\\
b&=\frac{||B_1-B_0||}{2}=\frac{n}{2}\sqrt{(\frac{b_{0x}}{b_{0z}}-\frac{b_{1x}}{b_{1z}})^2+(\frac{b_{0y}}{b_{0z}}-\frac{b_{1y}}{b_{1z}})^2}\\
\alpha &= acos(\vb{j} \cdot \vb{\hat{ME}})
\end{aligned}
\end{equation*}

To find the bounds of such an ellipse, we use a shear transformation to deform it into an axis-aligned while fixing a coordinate. By applying a vertical and horizontal shear transformation, we calculate the x and y bounds of the original ellipse. To make the transformation calculation easier, we first translate our ellipse center to the origin $O$. The equation of an ellipse with its center at the origin $O$, inclined at an angle $\alpha$, with $a$ its major radius and $b$ its shorter radius, is expressed as:
\begin{equation}
\frac{(x \cdot cos \alpha - y \cdot sin \alpha)^2}{a^2} + \frac{(x \cdot sin \alpha + y \cdot cos \alpha)^2}{b^2} = 1
\label{ellipse-eq}
\end{equation}

The shear transformation ensures that the area of the ellipse is preserved, hence we have:

\[\pi cd=\pi ab \Leftrightarrow c = \frac{ab}{d}\]

The distance $2d$ can be calculated by inputting $x=0$ in (\ref{ellipse-eq}). When $x=0$, we have $y=\pm d$, we  get rid of the sign because of the square exponent in the equation, thus we have:

\begin{equation*}
\begin{aligned}
&\qquad
\frac{(d \cdot sin \alpha)^2}{a^2} + \frac{(d \cdot cos \alpha)^2}{b^2} = 1\\
\Leftrightarrow&\qquad
\frac{d^2}{a^2b^2}(b^2\cdot sin^2 \alpha + a^2 \cdot cos^2 \alpha) = \frac{1}{c^2}(b^2\cdot sin^2 \alpha + a^2 \cdot cos^2 \alpha) = 1\\
\Leftrightarrow&\qquad
c=\pm\sqrt{b^2\cdot sin^2 \alpha + a^2 \cdot cos^2 \alpha}
\end{aligned}
\end{equation*}

Finally we translate our solution to our ellipse with center point E, giving us the bounds of the ellipse intersection:

\begin{equation}
\begin{aligned}
l'&=E_x - \sqrt{a^2 \cdot cos^2\alpha + b^2 \cdot sin^2\alpha}\\
r'&=E_x + \sqrt{a^2 \cdot cos^2\alpha + b^2 \cdot sin^2\alpha}
\end{aligned}
\label{x-min-max-ellipse}
\end{equation}

We do the same process with a horizontal shear transform that will give us:

\begin{equation}
\begin{aligned}
b'&=E_y - \sqrt{a^2 \cdot sin^2\alpha + b^2 \cdot cos^2\alpha}\\
t'&=E_y + \sqrt{a^2 \cdot sin^2\alpha + b^2 \cdot cos^2\alpha}
\end{aligned}
\label{y-min-max-ellipse}
\end{equation}


%
%
%
%
%
%
%
%

\bibliographystyle{abbrv-doi}

\bibliography{template}
\end{document}